\documentclass[a4paper, 11pt, twoside]{article}
\usepackage{latexsym,bm}
\usepackage[tbtags]{amsmath}
\usepackage{amssymb}

\usepackage{fancyhdr}
\usepackage{cite}
\usepackage{tabularx}
\usepackage{booktabs}
\usepackage{dcolumn}

\usepackage{multirow}

\usepackage{graphicx}

\usepackage{wrapfig}
\usepackage{multicol}
\usepackage{verbatim}
\usepackage[english]{babel}
\usepackage[T1]{fontenc}
\usepackage[utf8]{inputenc}
\usepackage{dsfont}
\usepackage{mathrsfs}
\usepackage{mathtools}
\usepackage{array}
\usepackage{hyperref}
\usepackage[scientific-notation=true]{siunitx}
\usepackage{xcolor}

\newcommand{\abs}[1]{\left\vert #1 \right\vert}

\newcommand{\vect}[1]{\boldsymbol{#1}}
\newcommand{\mtx}[1]{\boldsymbol{#1}}
\newcommand{\Tr}{\textrm{Tr}}
\newcommand{\grad}{\vect{\nabla}}
\newcommand{\dv}{{\rm div}}

\newcommand{\pj}{\mtx{P}}
\newcommand{\id}{\mtx{\mathrm{Id}}}
\newcommand{\rh}{\frac{r}{h}}
\newcommand{\rd}{{\rm d}}

\def\nkyear{0}                           
\def\nkvol{0}                              
\def\nkno{0}                                
\def\firstpage{1}                           
\def\DOI{0}
\def\shorttitle{Stable and accurate schemes for SDPD}
\def\shortauthor{G\'er\^ome Faure, Gabriel Stoltz}
\usepackage{Shangda1226}                    

\begin{document}
  
\title{{\large \textbf{Stable and accurate schemes for Smoothed Dissipative Particle Dynamics}}
}

\author{\small{G\'er\^ome Faure$^{1}$, \quad Gabriel Stoltz$^{2}$}
  \\[2mm]
  \footnotesize{1. CEA, DAM, DIF, F-91297 Arpajon, France;}
  \\
  \footnotesize{2. Universit\'e Paris-Est, CERMICS (ENPC), INRIA, F-77455 Marne-la-Vall\'ee, France }
}
  
\maketitle
\footnotesize

\begin{abstract}
  \noindent \textbf{Abstract~~~}  Smoothed Dissipative Particle Dynamics (SDPD) is a mesoscopic particle method which allows to select the level of resolution at which a fluid is simulated.
  The numerical integration of its equations of motion still suffers from the lack of numerical schemes satisfying all the desired properties such as energy conservation and stability.
  The similarities between SDPD and Dissipative Particle Dynamics with Energy conservation (DPDE), which is another coarse-grained model, enable the adaptation of recent numerical schemes developed for DPDE to the SDPD setting.
  In this article, we introduce a Metropolis step in the integration of the fluctuation/dissipation part of SDPD to improve its stability.
  \\[2mm]
  \textbf{Key words~~~} Smoothed Dissipative Particle Dynamics, Numerical integration, Metropolis algorithm
  \\[2mm]
  \textbf{2010 Mathematics Subject Classification~~~} 82-08, 65C30, 82C80
\end{abstract}

\section{Introduction}
\label{sec:introduction}
The development of the computational capacities in the last decades has allowed physicists to use numerical simulations to study physical properties at the atomic scale with the help of statistical physics.
In particular Molecular Dynamics (MD) consists in integrating the equation of motions for the atoms in order to sample probability measures in a high dimensional space~\cite{book_frenkel_2001,book_tuckerman_2010,book_leimkuhler_2015}.
However, traditional microscopic methods suffer from limitations in terms of accessible time and length scales, which drives the development of mesoscopic coarse-grained methods.
These mesoscopic models aim at greatly reducing the number of degrees of freedom explicitly described, and thus the computational cost, while retaining some properties absent from more macroscopic models such as hydrodynamics.
Smoothed Dissipative Particle Dynamics (SDPD)~\cite{espanol_2003} belongs to this class of mesoscopic method.
It couples a particle Lagrangian discretization of the Navier-Stokes, Smoothed Particle Hydrodynamics (SPH)~\cite{lucy_1977,monaghan_1977}, and the thermal fluctuations from models like Dissipative Particle Dynamics with Energy conservation (DPDE)~\cite{avalos_1997,espanol_1997}.
It is thus able to deal with hydrodynamics at nanoscale and has been shown to give results consistent with MD for a wide range of resolutions, at equilibrium and for shock waves~\cite{faure_2016}, or for dynamical properties such as the diffusion coefficient of a colloid in a SDPD bath~\cite{vazquez_2009,litvinov_2009}.
SDPD has in particular been used to study colloids~\cite{vazquez_2009,bian_2012}, polymer suspensions~\cite{litvinov_2008} and fluid mixtures~\cite{petsev_2016}.

One of the main challenges for mesoscopic models incorporating fluctuations is to develop efficient, stable and parallelizable numerical schemes for the integration of their stochastic dynamics.
Most schemes are based on a splitting strategy~\cite{trotter_1959,strang_1968} where the Hamiltonian part is integrated through a Velocity Verlet scheme~\cite{verlet_1967}.
A traditional and popular algorithm first proposed for Dissipative Particle Dynamics~\cite{hoogerbrugge_1992} and later extended to DPDE~\cite{stoltz_2006} relies on a pairwise treatment of the fluctuation/dissipation part~\cite{shardlow_2003}.
The adaptation of this scheme to dynamics preserving various invariants has led to a class of schemes called Shardlow-like Splitting Algorithms (SSA)~\cite{lisal_2011}.
A major drawback in this strategy is the complexity of its parallelization~\cite{larentzos_2014}.
Other schemes have been recently proposed in~\cite{homman_2016} to enhance its use in parallel simulations.

All these schemes are however hindered by instabilities when internal energies become negative.
This especially happens at small temperatures or when small heat capacities are considered, typically for small mesoparticles.
It has been proposed to use Monte Carlo principles sample the invariant measure of DPDE, by resampling the velocities along the lines of centers according to a Maxwell-Boltzmann distribution and redistributing the energy variation into internal energies according to some prescription~\cite{langenberg_2016}.
This approach leads however to a dynamics which is not consistent with DPDE.
It was proposed in~\cite{stoltz_2017} to correct discretization schemes for DPDE by rejecting unlikely or forbidden moves through a Metropolis procedure,  which prevents the appearance of negative internal energies and improves the stability of the integration schemes.

There exist relatively few references in the literature about the integration of the full SDPD dynamics.
Most work focus on numerical schemes in the isothermal setting~\cite{litvinov_2010}, avoiding the need to preserve the total energy during the simulation. 
In a previous article~\cite{faure_2016}, we introduced an adaptation of the Shardlow splitting to SDPD, allowing a good control of the energy conservation.
The aim of this work is to provide more details about the possible integration of SDPD in an energy conserving framework and most importantly to increase the stability for small particle sizes by adapting the Metropolization procedure described in~\cite{stoltz_2017}.

This article is organized as follows.
We first present in Section~\ref{sec:equations} the equations of SDPD as reformulated in~\cite{faure_2016}.
In Section~\ref{sec:schemes}, we recall the Shardlow splitting for SDPD and introduce a Metropolis step to enhance the stability of the algorithm.
We evaluate the properties of the Shardlow and Metropolis schemes by means of numerical simulations in Section~\ref{sec:results}.
Our conclusions are gathered in Section~\ref{sec:conclusion}.

\section{Smoothed Dissipative Particle Dynamics}
\label{sec:equations}
At the hydrodynamic scale, the dynamics of the fluid is governed by the Navier-Stokes equations~\eqref{eq:navier-stokes}, which read in their Lagrangian form when the heat conduction is neglected (for times $t\geq0$ and positions $\vect{x}$ in a domain $\Omega\subset \mathbb{R}^3$):
\begin{equation}
  \label{eq:navier-stokes}
  \begin{aligned}
    {\rm D}_t\rho + \rho\,\dv_{\vect{x}}\vect{v} &= 0,\\
    \rho {\rm D}_t\vect{v} &= \dv_{\vect{x}}\left(\mtx{\sigma}\right),\\
    \rho{\rm D}_t\left(u + \frac12\vect{v}^2\right) &= \dv_{\vect{x}}\left(\mtx{\sigma}\vect{v}\right).
  \end{aligned}
\end{equation}
In these equations, the material derivative used in the Lagrangian description is defined as
\[
  D_t f(t,\vect{x}) = \partial_t f(t,\vect{x}) + \vect{v}(t,\vect{x})\grad_{\vect{x}}f(t,\vect{x}).
\]
The unknowns are $\rho(t,\vect{x}) \in \mathbb{R}_+$ the density of the fluid, $\vect{v}(t,\vect{x}) \in \mathbb{R}^3$ its velocity, $u(t,\vect{x}) \in \mathbb{R}$ its internal energy and $\mtx{\sigma}(t,\vect{x}) \in \mathbb{R}^{3\times 3}$ the stress tensor:
\begin{equation}
\label{eq:stress-tensor}
  \mtx{\sigma} = P\id + \eta(\grad \vect{v} + (\grad\vect{v})^T) + \left(\zeta-\frac23\eta\right)\dv(\vect{v})\id,
\end{equation}
where $P$ is the pressure of the fluid, $\eta$ the shear viscosity and $\zeta$ the bulk viscosity.

In the following, we first present the SPH discretization of the Navier-Stokes equations in Section~\ref{sec:sph} before introducing the SDPD equations reformulated in terms of internal energies~\cite{faure_2016} in Section~\ref{sec:eom-sdpd}.

\subsection{Smoothed Particle Hydrodynamics}
\label{sec:sph}
Smoothed Particle Hydrodynamics~\cite{lucy_1977,monaghan_1977} is a Lagrangian discretization of the Navier-Stokes equations~(\ref{eq:navier-stokes}) on a finite number $N$ of fluid particles playing the role of interpolation nodes.
These fluid particles are associated with a portion of fluid of mass $m$.
They are located at positions $\vect{q}_i \in \Omega$ and have a momentum $\vect{p}_i \in\mathbb{R}^{3}$.
The internal degrees of freedom are represented by an internal energy $\varepsilon_i$.
In general, the energies are bounded below.
Upon shifting the minimum of the internal energy, we may consider that the internal energies remain positive ($\varepsilon_i>0$).

\subsubsection{Approximation of field variables and their gradients}
\label{sec:approx-sph}

A key ingredient in the SPH discretization is the use of a particle-based interpolation of the field variables.
This leads to an approximation of the field variables by averaging over their values at the particle positions weighted by a smoothing kernel function $W$.
The kernel is generally required to be non-negative, regular, normalized as $\int_{\Omega} W(\vect{r})d\vect{r} = 1$ and with finite support~\cite{book_liu_2003}.
We introduce the smoothing length $h$ defined such that $W(\vect{r})=0$ if $\abs{\vect{r}} \geq h $.
In the sequel, we use the notation $r = \abs{\vect{r}}$.
In this work, we rely on a cubic spline~\cite{liu_2003}, whose expression reads
\begin{equation}
  \label{eq:sdpd-cubic-w}
  W(\vect{r}) = \left\{
    \begin{array}{cl}
      \displaystyle \frac{8}{\pi h^3} \left(1-6\frac{r^2}{h^2}+6\frac{r^3}{h^3}\right) & \displaystyle \text{ if } r \leq \frac{h}{2},\\[1em]
      \displaystyle \frac{16}{\pi h^3} \left(1-\frac{r}{h}\right)^3 & \displaystyle \text{ if } \frac{h}{2} \leq r \leq h,\\[1em]
      0 & \displaystyle  \text{ if } r \geq h.
    \end{array}
  \right.
\end{equation}
The field variables are then approximated as
\begin{equation}
  \label{eq:sph-approx}
  f(\vect{x}) \approx \sum_{i=1}^N f_i W(\vect{x}-\vect{q}_i),
\end{equation}
where $f_i$ denotes the value of the field $f$ on the particle~$i$.

The approximation of the gradient $\grad_{\vect{x}} f$ is obtained by deriving equation~\eqref{eq:sph-approx}, which yields
\[
  \grad_{\vect{x}} f(\vect{x}) \approx  \sum_{i=1}^N f_i \grad_{\vect{x}}W(\abs{\vect{x}-\vect{q}_i}).
\]
In order to have more explicit expressions, we introduce the function $F$ such that $\vect{\nabla}_{\vect{r}} W(\vect{r}) = -F(\abs{\vect{r}})\vect{r}$.
In the case of the cubic spline~(\ref{eq:sdpd-cubic-w}), it reads
\[
  F(r) = \left\{
    \begin{array}{cl}
      \displaystyle \frac{48}{\pi h^5} \left(2-3\rh\right) & \displaystyle \text{ if } r \leq \frac{h}{2},\\[1em]
      \displaystyle \frac{48}{\pi h^5} \frac1{r} \left(1-\rh\right)^2& \displaystyle \text{ if } \frac{h}{2} \leq r \leq h,\\[1em]
      0 & \displaystyle \text{ if } r \geq h.
    \end{array}
    \right.
\]
The gradient approximation can then be rewritten as 
\[
  \grad_{\vect{x}} f(\vect{x}) \approx  -\sum_{i=1}^N f_i F(\abs{\vect{x}-\vect{q}_i})(\vect{x}-\vect{q}_i).
\]

In order to simplify the notation, we define the following quantities for two particles $i$ and $j$:
\[
  \vect{r}_{ij} = \vect{q}_i - \vect{q}_j,\quad
  r_{ij} = \abs{\vect{r}_{ij}},\quad
  \vect{e}_{ij} = \frac{\vect{r}_{ij}}{r_{ij}},\quad
  F_{ij} = F(r_{ij}).
\]
We can associate a density $\rho_i$ and volume $\mathcal{V}_i$ to each particle as
\begin{equation}
  \label{eq:sdpd-rho-v}
  \rho_i(\vect{q}) = \sum_{j=1}^N mW(\vect{r}_{ij}),\quad
  \mathcal{V}_i(\vect{q}) = \frac{m}{\rho_i(\vect{q})}.
\end{equation}
The corresponding approximations of the density gradient evaluated at the particle points read
\begin{equation}
  \label{eq:gradient-rho}
  \grad_{\vect{q}_j} \rho_i = \left\{
    \begin{array}{cl}
      m F_{ij}\vect{r}_{ij} & \text{ if } j\neq i,\\[.5em]
      -m \sum\limits_{j=1}^N F_{ij}\vect{r}_{ij} & \text{ if } j=i.
    \end{array}
  \right.
\end{equation}

\subsubsection{Thermodynamic closure}
\label{sec:thermo-closure}
As in the Navier-Stokes equations, an equation of state is required to close the set of equations provided by the SPH discretization.
This equation of state relates the entropy $S_i$ of the mesoparticle $i$ with its density $\rho_i(\vect{q})$ (as defined by~\eqref{eq:sdpd-rho-v}) and its internal energy $\varepsilon_i$ through an entropy function
\begin{equation}
  \label{eq:sdpd-eos}
  S_i(\varepsilon_i,\vect{q})=\mathcal{S}(\varepsilon_i,\rho_i(\vect{q})).
\end{equation}
The equation of state $\mathcal{S}$ can be computed by microscopic simulations or by an analytic expression modeling the material behavior.
It is then possible to assign to each particle a temperature
\[
  T_i(\varepsilon_i,\vect{q}) = \left[\frac1{\partial_{\varepsilon}\mathcal{S}}\right](\varepsilon_i,\rho(\vect{q})),
\]
pressure
\[
  P_i(\varepsilon_i,\vect{q}) = -\frac{\rho(\vect{q})^2}{m}\left[\frac{\partial_{\rho}\mathcal{S}}{\partial_{\varepsilon}\mathcal{S}}\right](\varepsilon_i,\rho(\vect{q})),
\]
and heat capacity at constant volume
\[
  C_i(\varepsilon_i,\vect{q}) = -\left[\frac{(\partial_{\varepsilon} \mathcal{S})^2}{\partial_{\varepsilon}^2\mathcal{S}}\right](\varepsilon_i,\rho(\vect{q})).
\]
To simplify the notation, we omit in Sections~\ref{sec:eom-sdpd} the dependence of $T_i$, $P_i$ and $C_i$ on the variables $\varepsilon_i$ and $\vect{q}$.

\subsection{Equations of motion for SDPD}
\label{sec:eom-sdpd}
Smoothed Dissipative Particle Dynamics~\cite{espanol_2003} is a top-down mesoscopic method relying on the SPH discretization of the Navier-Stokes equations with the addition of thermal fluctuations which are modeled by a stochastic force.
In its energy reformulation~\cite{faure_2016}, SDPD is a set of stochastic differential equations for the following variables: the positions $\vect{q}_i\in\Omega\subset\mathbb{R}^{3}$, the momenta $\vect{p}_i\in\mathbb{R}^{3}$ and the energies $\varepsilon_i\in \mathbb{R}$ for $i=1\dots N$.

The dynamics can be split into several elementary dynamics, the first one being a conservative dynamics derived from the pressure part of the stress tensor~\eqref{eq:stress-tensor} (see Section~\ref{sec:conservative-sdpd}) and a set of pairwise fluctuation and dissipation dynamics stemming from the viscous terms in~\eqref{eq:stress-tensor} coupled with random fluctuations (see Section~\ref{sec:fd-sdpd}).

\subsubsection{Conservative forces}
\label{sec:conservative-sdpd}
The elementary force between particles $i$ and $j$ arising from the discretization of the pressure gradient in the Navier-Stokes momentum equation reads
\begin{equation}
  \label{eq:cons-forces}
  \vect{\mathcal{F}}_{{\rm cons},ij} = m^2\left(\frac{P_i}{\rho_i^2}+\frac{P_j}{\rho_j^2}\right)F_{ij}\vect{r}_{ij}.
\end{equation}
In its original formulation~\cite{espanol_2003}, this conservative dynamics clearly appears as a Hamiltonian dynamics with a potential $\mathcal{E}(\vect{q},S_i)$ relating the energy with the positions and the particle entropy $S_i$.
The entropies are then invariants of this subdynamics.
In the energy reformulation, the entropies are no longer considered as such.
Instead the focus is on the total energy
\[
  E(\vect{q},\vect{p},\varepsilon) = \sum_{i=1}^N \varepsilon_i + \frac{\vect{p}_i^2}{2m},
\]
which is preserved by the dynamics.
This can be ensured by computing the variation of the particle volume as
\[
  \rd\mathcal{V}_i = -\frac{m}{\rho_i^2}\rd\rho_i = \sum_{j\neq i} \frac{m^2}{\rho_i^2}F_{ij}\vect{r}_{ij}\cdot\vect{v}_{ij}\rd t,
\]
leading to the variation of the internal energy given by
\[
  \rd\varepsilon_i = - P_i\rd\mathcal{V}_i+ T_i\rd S_i = -\sum_{j\neq i}\frac{m^2P_i}{\rho_i(\vect{q})^2}F_{ij}\vect{r}_{ij}\cdot\vect{v}_{ij}\,\rd t.
\]
This allows us to write the conservative part of the dynamics as
\begin{equation}
  \label{eq:sdpd-cons}
  \left\{
    \begin{aligned}
      \rd\vect{q}_i &= \frac{\vect{p}_i}{m}\,\rd t,\\
      \rd\vect{p}_i &= \sum_{j\neq i} \vect{\mathcal{F}}_{{\rm cons},ij}\,\rd t,\\
      \rd\varepsilon_i &= -\sum_{j\neq i}\frac{m^2P_i}{\rho_i(\vect{q})^2}F_{ij}\vect{r}_{ij}\cdot\vect{v}_{ij}\,\rd t.
    \end{aligned}
  \right.
\end{equation}
This dynamics preserve by construction the total momentum $\displaystyle \sum_{i=1}^N \vect{p}_i$ and the total energy $E(\vect{q},\vect{p},\varepsilon)$.

\subsubsection{Fluctuation and Dissipation}
\label{sec:fd-sdpd}
In order to give the expression of the viscous and fluctuating part of the dynamics, we define the relative velocity for a pair of particles $i$ and $j$ as
\[
\vect{v}_{ij} = \frac{\vect{p}_i}{m}-\frac{\vect{p}_j}{m}.
\]
The viscous term in the Navier-Stokes equations~(\ref{eq:navier-stokes}) is discretized by a pairwise dissipative force, while the thermal fluctuations are modeled by a pairwise stochastic force.
In the spirit of DPDE, the pairwise fluctuation and dissipation dynamics for $i<j$ is chosen of the following form:
\begin{equation}
  \label{eq:sdpd-simple-fluct}
  \left\{
  \begin{aligned}
    \rd\vect{p}_i &= -\mtx{\Gamma}_{ij}\vect{v}_{ij}\,\rd t + \mtx{\Sigma}_{ij}\rd\vect{B}_{ij},\\
    \rd\vect{p}_j &= \mtx{\Gamma}_{ij}\vect{v}_{ij}\,\rd t - \mtx{\Sigma}_{ij}\rd\vect{B}_{ij},\\
    \rd\varepsilon_i &= \frac12\left[\vect{v}_{ij}^T\mtx{\Gamma}_{ij}\vect{v}_{ij} - \frac{\Tr(\mtx{\Sigma}_{ij}\mtx{\Sigma}_{ij}^T)}{m}\right]\rd t -\frac12 \vect{v}_{ij}^T\mtx{\Sigma}_{ij}\rd\vect{B}_{ij},\\
    \rd\varepsilon_j &= \frac12\left[\vect{v}_{ij}^T\mtx{\Gamma}_{ij}\vect{v}_{ij} - \frac{\Tr(\mtx{\Sigma}_{ij}\mtx{\Sigma}_{ij}^T)}{m}\right]\rd t -\frac12 \vect{v}_{ij}^T\mtx{\Sigma}_{ij}\rd\vect{B}_{ij},
  \end{aligned}
  \right.
\end{equation}
where $\vect{B}_{ij}$ is a $3$-dimensional vector of standard Brownian motions and $\mtx{\Gamma}_{ij}$, $\mtx{\Sigma}_{ij}$ are $3\times3$ symmetric matrices.
by construction, \eqref{eq:sdpd-simple-fluct} preserves the total momentum in the system since $\rd \vect{p}_i = -\rd\vect{p}_j$
Furthermore, as in DPDE, the equations for the energy variables are determined to ensure the conservation of the total energy $E(\vect{q},\vect{p},\varepsilon)$.
As $\displaystyle \rd \varepsilon_i = -\frac12 \rd \left(\frac{\vect{p}_i^2}{2m} + \frac{\vect{p}_j^2}{2m}\right)$, It\^o calculus yields the resulting equations in~\eqref{eq:sdpd-simple-fluct}.

We consider friction and fluctuation matrices of the form
\begin{equation}
  \label{eq:fluct-gamma}
  \begin{aligned}
    \mtx{\Gamma}_{ij} &= \gamma^{\parallel}_{ij}(\varepsilon_i,\varepsilon_j,\vect{q})\pj^{\parallel}_{ij} + \gamma^{\perp}_{ij}(\varepsilon_i,\varepsilon_j,\vect{q})\pj^{\perp}_{ij},\\
    \mtx{\Sigma}_{ij} &= \sigma^{\parallel}_{ij}(\varepsilon_i,\varepsilon_j,\vect{q})\pj^{\parallel}_{ij} + \sigma^{\perp}_{ij}(\varepsilon_i,\varepsilon_j,\vect{q})\pj^{\perp}_{ij},
  \end{aligned}
\end{equation}
with the projection matrices $\pj^{\parallel}_{ij}$ and $\pj^{\perp}_{ij}$ given by
\[
  \pj_{ij}^{\parallel} = \vect{e}_{ij}\otimes\vect{e}_{ij},\quad
  \pj_{ij}^{\perp} = \id - \pj_{ij}^{\parallel}.
\]
Introducing the coefficients
\[
  \kappa_{ij}^{\parallel} = \left(\frac{10}3\eta+4\zeta\right)\frac{m^2F_{ij}}{\rho_i\rho_j},\quad \kappa_{ij}^{\perp} =\left(\frac{5\eta}{3}-\zeta\right)\frac{m^2F_{ij}}{\rho_i\rho_j},
\]
defined from the fluid viscosities $\eta$ and $\zeta$ appearing in the stress tensor~(\ref{eq:stress-tensor}), we can choose the friction and fluctuations coefficients as
\begin{equation}
  \label{eq:sdpd-gamma-sigma}
  \begin{aligned}
    d_{ij}(\varepsilon_i,\varepsilon_j,\vect{q}) &= k_{\rm B}\frac{T_iT_j}{(T_i+T_j)^2}\left(\frac1{C_i}+\frac1{C_j}\right),\\
    \gamma^{\theta}(\varepsilon_i,\varepsilon_j,\vect{q}) &= \kappa_{ij}^{\theta}\left( 1 - d_{ij}(\varepsilon_i,\varepsilon_j,\vect{q}) \right),\\
    \sigma^{\theta}(\varepsilon_i,\varepsilon_j,\vect{q}) &= 2\sqrt{\kappa_{ij}^{\theta} k_{\rm B}\frac{T_iT_j}{T_i+T_j}}.
  \end{aligned}
\end{equation}
As shown is~\cite{faure_2016}, this ensures that measures of the form (with $g$ a given smooth function)
\begin{equation}
  \label{eq:sdpd-energy-minv}
    \mu(\rd\vect{q}\,\rd\vect{p}\,\rd \varepsilon) = g\left(E(\vect{q},\vect{p},\varepsilon),\sum\limits_{i=1}^N\vect{p}_i\right)\prod_{i=1}^N\frac{\exp\left(\frac{S_i(\varepsilon_i,\vect{q})}{k_{\rm B}}\right)}{T_i(\varepsilon_i,\vect{q})}\,\rd\vect{q}\,\rd\vect{p}\,\rd \varepsilon
\end{equation}
are left invariant by the elementary dynamics~\eqref{eq:sdpd-simple-fluct}.
Alternative fluctuation/dissipation relations are possible (such as constant $\sigma$ parameters) but the relations~(\ref{eq:sdpd-gamma-sigma}) allow to retrieve the original SPDP~\cite{espanol_2003}.

\subsubsection{Complete equations of motion}
\label{sec:full-eom}
Gathering all the terms, the SDPD equations of motion reformulated in the position, momentum and internal energy variables read
\begin{equation}
  \label{eq:sdpd-energy}
  \left\{
  \begin{aligned}
    \rd\vect{q}_i =&\, \frac{\vect{p}_i}{m}\,\rd t,\\
    \rd\vect{p}_i =& \sum_{j\neq i} m^2\left(\frac{P_i}{\rho_i^2}+\frac{P_j}{\rho_j^2}\right)F_{ij}\vect{r}_{ij}\,\rd t - \mtx{\Gamma}_{ij}\vect{v}_{ij}\,\rd t + \mtx{\Sigma}_{ij}\rd\vect{B}_{ij},\\
    \rd \varepsilon_i =& \sum_{j\neq i} -\frac{m^2P_i}{\rho_i^2}F_{ij}\vect{r}_{ij}^T\vect{v}_{ij}\,\rd t + \frac12 \left[\vect{v}_{ij}^T\mtx{\Sigma}_{ij}\vect{v}_{ij} -\frac1{m}\Tr(\mtx{\Sigma}_{ij}\mtx{\Sigma}_{ij}^T)\right]\rd t
    - \frac12 \vect{v}_{ij}^T\mtx{\Sigma}_{ij}\rd\vect{B}_{ij},
  \end{aligned}
  \right.
\end{equation}
with $\mtx{\Sigma}_{ij}$ and $\mtx{\Gamma}_{ij}$ given by~\eqref{eq:fluct-gamma} and~(\ref{eq:sdpd-gamma-sigma}).
The dynamics~\eqref{eq:sdpd-energy} preserves the total momentum $\sum\limits_{i=1}^N\vect{p}_i$ and the total energy $E(\vect{q},\vect{p},\varepsilon)$ since all the elementary subdynamics ensure these conservations.

\subsection{Reduced units for SDPD}
\label{sec:scaling-sdpd}

In SDPD, the mass $m$ of the fluid particles allows us to change the resolution of the method.
We introduce the particle size $\displaystyle K = \frac{m}{m_0}$, where $m_0$ is the mass of one microscopic particle (\emph{e.g.} a molecule).
Since we deal with different particle sizes in the following, it is convenient to introduce reduced units for each size $K$:
\begin{equation}
  \label{eq:reduced-units}
  \begin{aligned}
    \widetilde{m}_K &= Km_0,\\
    \widetilde{l}_K &= \left(\frac{Km_0}{\rho}\right)^{\frac13},\\
    \widetilde{\varepsilon}_K &= K k_{\rm B}T,
  \end{aligned}
\end{equation}
where $\widetilde{m}_K$ is the mass unit, $\widetilde{l}_K$ the length unit, $\widetilde{\varepsilon}_K$ the energy unit and $\rho$ the average density of the fluid.
With such a set of reduced units, the time unit is 
\[
  \widetilde{t}_K = \widetilde{l}_K\sqrt{\frac{\widetilde{m}_K}{\widetilde{\varepsilon}_K}} = \frac{m_0^{\frac56}K^{\frac13}}{\rho^{\frac13} \sqrt{k_{\rm B}T}}.
\]
In the following, we select time steps before expressing them in terms of $\widetilde{t}_K$, with $K$ the particle size used in the simulations.
This explains the use of non round time steps in Section~\ref{sec:results}.

The smoothing length $h_K$ defining the cut-off radius in~\eqref{eq:sdpd-cubic-w} also needs to be adapted to the size of the SDPD particles so that the approximations~(\ref{eq:sph-approx}) continue to make sense.
In order to keep the average number of neighbors roughly constant in the smoothing sum, $h_K$ should be rescaled as
\[
  h_K = h\left(\frac{m_K}{\rho}\right)^{\frac13}.
\]
In this work, we take $h=2.5$, which corresponds to a typical number of 60-70 neighbors, a commonly accepted number~\cite{liu_2003}.

\section{Integration schemes}
\label{sec:schemes}
In the following, we describe several numerical schemes for the integration of SDPD.
They all rely on a splitting strategy~\cite{trotter_1959,strang_1968} where the full dynamics is divided in simpler elementary dynamics that are consecutively integrated.
Since the conservative part of the dynamics~(\ref{eq:sdpd-cons}) can be viewed as a Hamiltonian dynamics, it is natural to resort to a symplectic scheme such as the widely used Velocity-Verlet scheme~\cite{verlet_1967} which ensures a good energy conservation in the long term~\cite{hairer_2003,book_hairer_2002}.
This algorithm is briefly described in Section~\ref{sec:verlet}.

There is however no definite way to deal with the fluctuation/dissipation part described in Section~\ref{sec:fd-sdpd}.
Due to its close similarities with DPDE, we propose in the following to adapt some schemes devoted to the integration of DPDE to the SDPD setting.
One approach to integrate SDPD, described in~\cite{faure_2016}, is based on the algorithm proposed by Shardlow~\cite{shardlow_2003} for DPD and its subsequent adaptations to DPDE~\cite{stoltz_2006}.
The dynamics is split into a Hamiltonian part, discretized through a Velocity-Verlet algorithm~(\ref{eq:sdpd-verlet}), and elementary pairwise fluctuation/dissipation dynamics that are successively integrated.
We first recall in Section~\ref{sec:ssa} the Shardlow-like splitting scheme (SSA) used in~\cite{faure_2016}.
While this provides a way to integrate SDPD preserving its invariants (approximately for the energy), it suffers from stability issues especially for small particle sizes, when the internal and kinetic energy are of the same scale.
We thus explore methods to improve the stability of these integration algorithms in Section~\ref{sec:metropolis}, relying on the ideas developed in~\cite{stoltz_2017} where a Metropolis acceptance-rejection step is included to correct the biases of the numerical discretization of the fluctuation/dissipation part.

\subsection{Integrating the Hamiltonian part of the dynamics}
\label{sec:verlet}

It is convenient to consider the conservative part of the dynamics~(\ref{eq:sdpd-cons}) in its original formulation in the position, momentum and entropy variables~\cite{espanol_2003} in order to take advantage of the conservation of the entropies $S_i$.
The internal energies are related to the positions and entropies by an energy function $\mathcal{E}(S_i,\rho_i(\vect{q}))$, which allows us to write the Hamiltonian as
\[
  H(\vect{q},\vect{p},S) = \sum_{i=1}^N \frac{\vect{p}_i^2}{2m} + \mathcal{E}(S_i,\rho_i(\vect{q})).
\]
The dynamics~(\ref{eq:sdpd-cons}) can thus be recast in Hamiltonian form as
\[
  \left\{
    \begin{aligned}
      \rd\vect{q}_i &= \frac{\vect{p}_i}{m}\,\rd t,\\
      \rd\vect{p}_i &= -\grad_{\vect{q}_i} \mathcal{H}(\vect{q},\vect{p},S) = \sum_{j\neq i} \vect{\mathcal{F}}_{{\rm cons},ij}\,\rd t.
    \end{aligned}
  \right.
\]
The Velocity-Verlet scheme~\cite{verlet_1967} allows to integrate such dynamics while preserving on average the Hamiltonian $H$.
This corresponds to the following integration scheme:
\begin{equation}
  \label{eq:sdpd-verlet}
  \left\{
  \begin{aligned}
    \vect{p}^{n+\frac12}_i &= \vect{p}^n_i + \sum_{j\neq i}\vect{\mathcal{F}}_{{\rm cons},ij}^n \frac{\Delta t}{2},\\
    \vect{q}^{n+1}_i &= \vect{q}^n_i + \frac{\vect{p}^{n+\frac12}_i}{m}\Delta t,\\
    \vect{p}^{n+1}_i &= \vect{p}^{n+\frac12}_i + \sum_{j\neq i}\vect{\mathcal{F}}_{{\rm cons},ij}^{n+1}\frac{\Delta t}{2}.
  \end{aligned}
  \right.
\end{equation}

\subsection{Shardlow-like Splitting Algorithm}
\label{sec:ssa}

We present here a first possibility for the integration of the fluctuation/dissipation dynamics introduced in~\cite{faure_2016} based on existing schemes for DPD~\cite{shardlow_2003} and DPDE~\cite{stoltz_2006}.
If we neglect the dependence of $\Gamma$ and $\Sigma$ on $\varepsilon_i$, the elementary dynamics~(\ref{eq:sdpd-simple-fluct}) on the momenta can be viewed as a standard Ornstein-Uhlenbeck process and solved analytically.
We provided in~\cite{faure_2016} the corresponding expression for the updated momenta after a time step $\Delta t$ as
\begin{equation}
  \label{eq:ssa-ou}
  \begin{pmatrix}
    \vect{p}_i^{n+1}\\[.5em]
    \vect{p}_j^{n+1}
  \end{pmatrix}
  = \sum\limits_{\theta\in\{\parallel,\perp\}} \pj_{ij}^{\theta}\left[\frac{m}2 \alpha_{ij}^{\theta}(\varepsilon_i^n,\varepsilon_j^n,\vect{q}^n) \vect{v}_{ij}^n + \zeta_{ij}^{\theta}(\varepsilon_i^n,\varepsilon_j^n,\vect{q}^n) \vect{G}_{ij}^n \right]
  \begin{pmatrix}
    \vect{1}\\[.5em]
    \vect{-1}
  \end{pmatrix},
\end{equation}
where $\vect{G}_{ij}^n$ is a standard 3-dimensional Gaussian variable and for $\theta \in \{\parallel,\perp\}$,
\[
  \begin{aligned}
    \alpha_{ij}^{\theta}(\varepsilon_i,\varepsilon_j,\vect{q}) &= \exp\left(-\frac{2\gamma_{ij}^{\theta}(\varepsilon_i,\varepsilon_j,\vect{q})\Delta t}{m}\right),\\
    \zeta_{ij}^{\theta}(\varepsilon_i,\varepsilon_j,\vect{q}) &= \sigma_{ij}^{\theta}(\varepsilon_i,\varepsilon_j,\vect{q})\sqrt{\frac{m(1-(\alpha_{ij}^{\theta}(\varepsilon_i,\varepsilon_j,\vect{q}))^2) }{4\gamma_{ij}^{\theta}(\varepsilon_i,\varepsilon_j,\vect{q})}}.
  \end{aligned}
\]
The integration of the momenta with~(\ref{eq:ssa-ou}) induces a variation of the kinetic energy which is then redistributed symmetrically in the internal energies as suggested in~\cite{marsh_1998,stoltz_2006}.
This guarantees the exact conservation of the energy during this elementary step.
The new internal energies are finally given by
\[
  \left\{
    \begin{aligned}
      \varepsilon_i^{n+1} &= \varepsilon_i^n -\frac12\left[\frac{\left(\vect{p}_i^{n+1}\right)^2}{2m} + \frac{\left(\vect{p}_j^{n+1}\right)^2}{2m} - \frac{\left(\vect{p}_i^n\right)^2}{2m} - \frac{\left(\vect{p}_j^n\right)^2}{2m}\right],\\
      \varepsilon_j^{n+1} &= \varepsilon_j^n -\frac12\left[\frac{\left(\vect{p}_i^{n+1}\right)^2}{2m} + \frac{\left(\vect{p}_j^{n+1}\right)^2}{2m} - \frac{\left(\vect{p}_i^n\right)^2}{2m} - \frac{\left(\vect{p}_j^n\right)^2}{2m}\right].
    \end{aligned}
  \right.
\]
Thermodynamic variables like the temperatures $T_i$, $T_j$ and heat capacities $C_i$, $C_j$ are updated with the equation of state using the new internal energies, before turning to another pair of particles.

Let us however remark that the pairwise Shardlow-like algorithm is sequential by nature and its parallelization requires a convoluted method~\cite{larentzos_2014}.
Moreover, and maybe more importantly, there is no mechanism preventing the apparition of negative energies during the simulation.
This situation happens when the fluctuations are large with respect to the internal energies: typically at low temperature or when the particle sizes are small (so that their heat capacity are small as well).
This leads to stability issues unless very small timesteps are used.

\subsection{Metropolized integration scheme}
\label{sec:metropolis}

To avoid instabilities related to negative internal energies while keeping reasonable time steps, it has been proposed to include a Metropolis step to reject impossible or unlikely moves~\cite{stoltz_2017}.
In the following, we show how this procedure can be used for SDPD.
First, we reformulate the pairwise dynamics~(\ref{eq:sdpd-simple-fluct}) as an overdamped Langevin dynamics in the relative velocity $\vect{v}_{ij}$ variable only, see Section~\ref{sec:fd-overdamped}.
We then construct proposed moves for the Metropolized scheme and compute the corresponding acceptance ratio in Section~\ref{sec:ratio}.
A simplified version of the Metropolized scheme is introduced in Section~\ref{sec:approx-met} where the computation of the Metropolis ratio is avoided and the rejection occurs only to avoid negative internal energies.

\subsubsection{Reformulation of the fluctuation and dissipation dynamics as an overdamped Langevin dynamics}
\label{sec:fd-overdamped}

In order to simplify the Metropolization of the integration scheme, we show that the elementary fluctuation-dissipation dynamics~(\ref{eq:sdpd-simple-fluct}) can be described only in terms of the relative velocity $\vect{v}_{ij}$ and formulated as an overdamped Langevin dynamics.

Since the dynamics~(\ref{eq:sdpd-simple-fluct}) preserve the momentum $\vect{p}_i+\vect{p}_j$, the momenta $\vect{p}_i$ and $\vect{p}_j$ can be rewritten as a function of $\vect{v}_{ij} $ as:
\[
  \vect{p}_i = \frac{\vect{p}_i+\vect{p}_j}2 + \frac{\vect{p}_i-\vect{p}_j}{2} = \frac{\vect{p}_i^0 + \vect{p}_j^0}2 + \frac{m}2 \vect{v}_{ij} = \vect{p}_i^0 + \frac{m}2(\vect{v}_{ij} - \vect{v}_{ij}^0),
\]
and
\[
  \vect{p}_j = \vect{p}_j^0 - \frac{m}2(\vect{v}_{ij} - \vect{v}_{ij}^0).
\]
This already shows how to express the momenta $\vect{p}_i$ and $\vect{p}_j$ in terms of $\vect{v}_{ij}$.
In addition, the kinetic energy formulated in the relative velocity reads
\[
  \frac{\vect{p}_i^2+\vect{p}_j^2}{2m} = \frac{\left(\vect{p}_i^0\right)^2+\left(\vect{p}_j^0\right)^2}{2m} + \frac{m}4(\vect{v}_{ij} - \vect{v}_{ij}^0)^2.
\]
The conservation of the energy $\frac{\vect{p}_i^2+\vect{p}_j^2}{2m}+\varepsilon_i+\varepsilon_j$ and the fact that $\rd \varepsilon_i = \rd \varepsilon_j$ provides the expression of the internal energies as a function of the relative velocity as
\begin{equation}
  \label{eq:energies-vij}
  \varepsilon_i = \varepsilon_i^0 - \frac{m}8\left(\left[\vect{v}_{ij}\right]^2 - \left[\vect{v}_{ij}^0\right]^2\right),\quad    \varepsilon_j = \varepsilon_j^0 - \frac{m}8\left(\left[\vect{v}_{ij}\right]^2 - \left[\vect{v}_{ij}^0\right]^2\right).
\end{equation}

Using this relation, the dynamics~(\ref{eq:sdpd-simple-fluct}) can in fact be rewritten as an effective dynamics on the relative velocity only, as
\begin{equation}
  \label{eq:fd-effective}
  \rd \vect{v}_{ij} = -\frac2m\mtx{\Gamma}_{ij}\vect{v}_{ij} \rd t + \frac2m\mtx{\Sigma}_{ij}\rd\vect{B}_{ij},
\end{equation}
where $\Gamma_{ij}$, $\Sigma_{ij}$ are functions of the relative velocity through~\eqref{eq:energies-vij}.
We claim that the dynamics~(\ref{eq:fd-effective}) can be written more explicitly as an overdamped Langevin dynamics under the form
\begin{equation}
  \label{eq:fd-overdamped}
  \rd \vect{v}_{ij} = \left(-\mtx{M}(\vect{v}_{ij})\grad_{\vect{v}_{ij}}\mathcal{U}(\vect{v}_{ij}) + \dv_{\vect{v}_{ij}}(\mtx{M})(\vect{v}_{ij})\right) \rd t + \sqrt2\mtx{M}^{\frac12}(\vect{v}_{ij})\rd\vect{B}_{ij},
\end{equation}
with the diffusion matrix
\[
  \mtx{M}(\vect{v}_{ij}) = \frac{2\left[\sigma^{\parallel}(\varepsilon_i,\varepsilon_j,\vect{q})\right]^2}{m^2}\pj_{ij}^{\parallel} + \frac{2\left[\sigma^{\perp}(\varepsilon_i,\varepsilon_j,\vect{q}))\right]^2}{m^2}\pj_{ij}^{\perp},
\]
and the potential
\[
  \mathcal{U}(\vect{v}_{ij}) = U\left(\varepsilon_i^0- \frac{m}{8}\left(\left[\vect{v}_{ij}\right]^2-\left[\vect{v}_{ij}^0\right]^2\right),\vect{q}\right) + U\left(\varepsilon_j^0- \frac{m}{8}\left(\left[\vect{v}_{ij}\right]^2-\left[\vect{v}_{ij}^0\right]^2\right),\vect{q}\right),
\]
where
\[
  U(\varepsilon_i,\vect{q}) = \log T_i(\varepsilon,i,\vect{q}) - \frac1{k_{\rm B}}S_i(\varepsilon_i,\vect{q}).
\]

Let us emphasize that the reformulation~\eqref{eq:fd-overdamped} is the key element for the Metropolis stabilization.
We now check that~\eqref{eq:fd-overdamped} holds.
By definition
\[
  \mtx{M}^{\frac12}(\vect{v}_{ij}) = \frac{\sqrt{2}}{m}\mtx{\Sigma}_{ij}.
\]
It therefore suffices to check that
\[
  -\frac2m\mtx{\Gamma}_{ij}\vect{v}_{ij} = -\mtx{M}(\vect{v}_{ij})\grad_{\vect{v}_{ij}}\mathcal{U}(\vect{v}_{ij}) + \dv_{\vect{v}_{ij}}(\mtx{M})(\vect{v}_{ij}).
\]
We first compute the gradient of the potential $\mathcal{U}$:
\[
  \begin{aligned}
    \grad_{\vect{v}_{ij}}\mathcal{U}(\vect{v}_{ij}) &= \left(\grad_{\vect{v}_{ij}} \varepsilon_i\right) \partial_{\varepsilon}U(\varepsilon_i,\vect{q}) + \left(\grad_{\vect{v}_{ij}} \varepsilon_j\right) \partial_{\varepsilon}U(\varepsilon_j,\vect{q})\\
    &= -\frac{m}4\vect{v}_{ij} \left( \frac{\partial_{\varepsilon_i}T_i}{T_i} - \frac{\partial_{\varepsilon_i}S_i}{k_{\rm B}} + \frac{ \partial_{\varepsilon_j}T_j}{T_j} - \frac{\partial_{\varepsilon_j} S_j}{ k_{\rm B}}  \right)\\
    &= -\frac{m}4\vect{v}_{ij} \left( \frac1{T_i}\left[\frac1{C_i}-\frac1{k_{\rm B}}\right] + \frac1{T_j}\left[\frac1{C_j}-\frac1{k_{\rm B}}\right] \right).
  \end{aligned}
\]
Upon application of the matrix $\mtx{M}$,
\[
  \begin{aligned}
    \mtx{M}(\vect{v}_{ij})\grad_{\vect{v}_{ij}}\mathcal{U}(\vect{v}_{ij})
    &= -\frac1{2m}\vect{v}_{ij} \sum_{\theta\in\{\parallel,\perp\}} \sigma^{\theta}(\varepsilon_i,\varepsilon_j,\vect{q})^2\left( \frac1{T_i}\left[\frac1{C_i}-\frac1{k_{\rm B}}\right] + \frac1{T_j}\left[\frac1{C_j}-\frac1{k_{\rm B}}\right] \right) \pj_{ij}^{\theta}\vect{v}_{ij}\\
    &= \frac2{m}\vect{v}_{ij} \sum_{\theta\in\{\parallel,\perp\}} \kappa_{ij}^{\theta}\left(1 - \frac{k_{\rm B}}{T_i+T_j}\left(\frac{T_j}{C_i} + \frac{T_i}{C_j}\right) \right) \pj_{ij}^{\theta}\vect{v}_{ij}.
  \end{aligned}
\]
The divergence of $\mtx{M}$ with respect to the relative velocity reads
\[
  \begin{aligned}
    \dv_{\vect{v}_{ij}}(\mtx{M})(\vect{v}_{ij}) &= \frac2{m^2}\pj_{ij}^{\parallel} \grad_{\vect{v}_{ij}}\left(\left[\sigma^{\parallel}(\varepsilon_i,\varepsilon_j,\vect{q})\right]^2\right) + \frac2{m^2}\pj_{ij}^{\perp} \grad_{\vect{v}_{ij}}\left(\left[\sigma^{\perp}(\varepsilon_i,\varepsilon_j,\vect{q})\right]^2\right)\\
    &= -\frac1{2m} \sum_{\theta\in\{\parallel,\perp\}} \left(\partial_{\varepsilon_i}+\partial_{\varepsilon_j}\right) \left(\left[\sigma^{\theta}(\varepsilon_i,\varepsilon_j,\vect{q})\right]^2\right) \pj_{ij}^{\theta}\vect{v}_{ij}\\
    &= -\frac2{m} k_{\rm B}
    \sum_{\theta\in\{\parallel,\perp\}} \kappa_{ij}^{\theta} \left(
      \frac{T_j(\partial_{\varepsilon_i}T_i)}{T_i+T_j} - \frac{T_iT_j(\partial_{\varepsilon_i} T_i)}{(T_i+T_j)^2}
      + \frac{T_i(\partial_{\varepsilon_j}T_j)}{T_i+T_j} - \frac{T_iT_j(\partial_{\varepsilon_j} T_j)}{(T_i+T_j)^2}
    \right)
    \pj_{ij}^{\theta}\vect{v}_{ij}\\
    &= \frac2{m}
    \sum_{\theta\in\{\parallel,\perp\}} \kappa_{ij}^{\theta} \left(
      d_{ij}
      - \frac{k_{\rm B}}{T_i+T_j}\left[\frac{T_j}{C_i}+\frac{T_i}{C_j}\right]
    \right)
    \pj_{ij}^{\theta}\vect{v}_{ij}.
  \end{aligned}
\]
The desired result follows from
\[
  \begin{aligned}
    -\mtx{M}(\vect{v}_{ij})\grad_{\vect{v}_{ij}}\mathcal{U}(\vect{v}_{ij}) + \dv_{\vect{v}_{ij}}(\mtx{M})(\vect{v}_{ij}) &= -\frac2m \left( (1-d_{ij})\kappa_{ij}^{\parallel}\pj_{ij}^{\parallel} + (1-d_{ij})\kappa_{ij}^{\perp}\pj_{ij}^{\perp}\right) \vect{v}_{ij}\\
    &= -\frac2m\mtx{\Gamma}_{ij}\vect{v}_{ij}.
  \end{aligned}
\]

In view of~\eqref{eq:fd-overdamped}, it can be immediately deduced that the following measure on the relative velocity, at fixed momenta $\vect{p}_i$, $\vect{p}_j$ and fixed internal energies $\varepsilon_i$, $\varepsilon_j$, is left invariant by the overdamped Langevin dynamics~(\ref{eq:fd-effective}):
\begin{equation}
  \label{eq:minv-vij}
  \nu(d \vect{v}_{ij}) = Z_{ij}^{-1}\exp\left(-\mathcal{U}(\vect{v}_{ij})\right)\rd \vect{v}_{ij} 
  = Z_{ij}^{-1}\frac{
    \exp\left(
      k_{\rm B}^{-1}
      \left[
        S_i\left(\varepsilon_i,\vect{q}\right)
        +
        S_j\left(\varepsilon_j,\vect{q}\right)
      \right]
    \right)
  }{
    T_i\left(\varepsilon_i,\vect{q}\right)
    T_j\left(\varepsilon_j,\vect{q}\right)
  }
  \rd \vect{v}_{ij}
\end{equation}

\subsubsection{Metropolis ratio}
\label{sec:ratio}

We consider~(\ref{eq:ssa-ou}) as the proposed move.
In terms of the relative velocity, it reads
\begin{equation}
  \label{eq:proposal}
  \begin{aligned}
    \vect{v}_{ij}^{n+1} &=
    \sum_{\theta\in\{\parallel,\perp\}}
    \alpha_{ij}^{\theta}(\varepsilon_i^n,\varepsilon_j^n,\vect{q}^n)\pj_{ij}^{\theta}\vect{v}_{ij}^n
    + \zeta_{ij}^{\theta}(\varepsilon_i^n,\varepsilon_j^n,\vect{q}^n)\pj_{ij}^{\theta}\vect{G}^n\\
    &= \mtx{\mathcal{A}}(\vect{q}^n,\vect{v}_{ij}^n)\vect{v}_{ij}^n + \mtx{\mathcal{B}}(\vect{q}^n,\vect{v}_{ij}^n)\vect{G}_{ij}^n,
  \end{aligned}
\end{equation}
with
\[
  \mtx{\mathcal{A}}(\vect{q}^n,\vect{v}_{ij}^n) =
  \alpha_{ij}^{\parallel}(\varepsilon_i^n,\varepsilon_j^n,\vect{q}^n)\pj_{ij}^{\parallel}
  +
  \alpha_{ij}^{\perp}(\varepsilon_i^n,\varepsilon_j^n,\vect{q}^n)\pj_{ij}^{\perp},
\]
and
\[
  \mtx{\mathcal{B}}(\vect{q}^n,\vect{v}_{ij}^n) =
  \zeta_{ij}^{\parallel}(\varepsilon_i^n,\varepsilon_j^n,\vect{q}^n)\pj_{ij}^{\parallel}
  +
  \zeta_{ij}^{\perp}(\varepsilon_i^n,\varepsilon_j^n,\vect{q}^n)\pj_{ij}^{\perp},
\]
The momenta and internal energies can then be updated as
\begin{equation}
  \label{eq:proposal-update}
  \left\{
  \begin{aligned}
    \vect{p}_i^{n+1} &= \vect{p}_i^n + \frac{m}{2}(\vect{v}_{ij}^{n+1}-\vect{v}_{ij}^n),\\
    \vect{p}_j^{n+1} &= \vect{p}_j^n - \frac{m}{2}(\vect{v}_{ij}^{n+1}-\vect{v}_{ij}^n),\\
    \varepsilon_i^{n+1} &= \varepsilon_i^n - \frac{m}{8}\left(\left[\vect{v}_{ij}^{n+1}\right]^2-\left[\vect{v}_{ij}^n\right]^2\right),\\
    \varepsilon_j^{n+1} &= \varepsilon_j^n - \frac{m}{8}\left(\left[\vect{v}_{ij}^{n+1}\right]^2-\left[\vect{v}_{ij}^n\right]^2\right).
  \end{aligned}
  \right.
\end{equation}
The internal energies $T_i$, $T_j$ and heat capacities $C_i$, $C_j$ are updated accordingly.

In order to decide whether we update the configuration with the proposed move or keep the current one, we first check whether $\varepsilon_i^{n+1}$ and $\varepsilon_j^{n+1}$ are negative, in which case the proposal is rejected.
Otherwise, we compute a Metropolis ratio that is an acceptance probability.
The probability to accept the proposed move from $\vect{v}$ to $\vect{v}'$ is $\min(1,A_{\Delta t}(\vect{v},\vect{v}'))$ with
\[
  A_{\Delta t}(\vect{v},\vect{v}') = \frac{\nu(\vect{v}')T_{\Delta t}(\vect{v}',\vect{v})}{\nu(\vect{v})T_{\Delta t}(\vect{v},\vect{v}')},
\]
where $T_{\Delta t}$ is the transition kernel associated with the proposal.
In the following, we omit all the dependence on the positions $\vect{q}^n$, which remain constant in this subdynamics, to simplify the notation.

The probability that~(\ref{eq:proposal}) proposes $\vect{v}'$ starting from $\vect{v}$ is given by
\[
  T_{\Delta t}(\vect{v},\vect{v}') =
  \frac1{(2\pi)^{\frac32}\abs{\mtx{\mathcal{B}}(\vect{v})}}
  \exp\left(-\frac12(\vect{v}'-\mtx{\mathcal{A}}(\vect{v})\vect{v})^T\mtx{\mathcal{B}}(\vect{v})^{-2}(\vect{v}'-\mtx{\mathcal{A}}(\vect{v})\vect{v})\right),
\]
with the inverse matrix $\displaystyle \mtx{\mathcal{B}}(\vect{v})^{-1} = \frac1{\zeta_{ij}^{\parallel}(\vect{v})}\pj_{ij}^{\parallel} + \frac1{\zeta_{ij}^{\perp}(\vect{v})}\pj_{ij}^{\perp}$ and the determinant $\abs{\mtx{\mathcal{B}}(\vect{v})} = \zeta_{ij}^{\parallel}(\vect{v})\zeta_{ij}^{\perp}(\vect{v})^2$.
For the direct move, the transition probability simply reads
\[
  T_{\Delta t}(\vect{v}_{ij}^n,\vect{v}_{ij}^{n+1}) =
  \frac1{(2\pi)^{\frac32}\abs{\mtx{\mathcal{B}}(\vect{v}_{ij}^n)}}
  \exp\left(-\frac{\left(\vect{G}^n\right)^T\vect{G}^n}2\right),
\]
while, for the reverse move,
\[
  T_{\Delta t}(\vect{v}_{ij}^{n+1},\vect{v}_{ij}^n) =
  \frac1{(2\pi)^{\frac32}\abs{\mtx{\mathcal{B}}(\vect{v}_{ij}^{n+1})}}
  \exp\left(-\frac12(\vect{v}_{ij}^n-\mtx{\mathcal{A}}(\vect{v}_{ij}^{n+1})\vect{v}_{ij}^{n+1})^T\mtx{\mathcal{B}}(\vect{v}_{ij}^{n+1})^{-2}(\vect{v}_{ij}^n-\mtx{\mathcal{A}}(\vect{v}_{ij}^{n+1})\vect{v}_{ij}^{n+1})\right).
\]
Using~(\ref{eq:minv-vij}) with the reference taken at iteration $n$,
\[
  \begin{aligned}
    \log\left(\frac{\nu(\vect{v}_{ij}^{n+1})}{\nu(\vect{v}_{ij}^n)}\right) =& \sum_{k\in\{i,j\}} \frac1{k_{\rm B}}\left[S_k\left(\varepsilon_k^n- \frac{m}{8}\left(\left[\vect{v}_{ij}^{n+1}\right]^2-\left[\vect{v}_{ij}^n\right]^2\right)\right) - S_k(\varepsilon_k^n)\right]\\
    &- \log\left[T_k\left(\varepsilon_k^n - \frac{m}{8}\left(\left[\vect{v}_{ij}^{n+1}\right]^2-\left[\vect{v}_{ij}^n\right]^2\right)\right)\right] + \log[T_k(\varepsilon_k^n)].
  \end{aligned}
\]
Finally, the acceptance ratio is given by
\begin{equation}
  \label{eq:metropolis-ratio}
  A_{\Delta t}(\vect{v}_{ij}^n,\vect{v}_{ij}^{n+1}) = \exp(a(\vect{v}_{ij}^n,\vect{v}_{ij}^{n+1})),
\end{equation}
with
\[
  \begin{aligned}
    a(\vect{v}_{ij}^n,\vect{v}_{ij}^{n+1}) &= -\log\left(T_{\Delta t}(\vect{v}_{ij}^n,\vect{v}_{ij}^{n+1})\right) + \log\left(T_{\Delta t}(\vect{v}_{ij}^{n+1},\vect{v}_{ij}^n)\right) + \log\left(\frac{\nu(\vect{v}_{ij}^{n+1})}{\nu(\vect{v}_{ij}^n)}\right)\\
    &= \frac{\left(\vect{G}^n\right)^T\vect{G}^n}2 + \log\abs{\mtx{\mathcal{B}}(\vect{v}_{ij}^n)}
     - \frac12(\vect{v}_{ij}^n-\mtx{\mathcal{A}}(\vect{v}_{ij}^{n+1})\vect{v}_{ij}^{n+1})^T\mtx{\mathcal{B}}(\vect{v}_{ij}^{n+1})^{-2}(\vect{v}_{ij}^n-\mtx{\mathcal{A}}(\vect{v}_{ij}^{n+1})\vect{v}_{ij}^{n+1})\\
    &\quad\,- \log\abs{\mtx{\mathcal{B}}(\vect{v}_{ij}^{n+1})} + \frac1{k_{\rm B}}\left(S_i(\varepsilon_i^{n+1}) - S_i(\varepsilon_i^n) + S_j(\varepsilon_j^{n+1}) - S_j(\varepsilon_j^n)\right)\\
    &\quad\,- \log\left(T_i(\varepsilon_i^{n+1})\right) + \log(T_i(\varepsilon_i^n)) - \log\left(T_j(\varepsilon_j^{n+1})\right) + \log(T_j(\varepsilon_j^n)).
  \end{aligned}
\]

Starting from a configuration $(\vect{p}_i^n,\vect{p}_j^n,\varepsilon_i^n,\varepsilon_j^n)$, the overall algorithm (Exact Metropolis Scheme or EMS) to integrate the fluctuation/dissipation for a pair $(i,j)$ of particle is organized as follows:
\begin{enumerate}
\item Compute a proposed move for $\vect{v}_{ij}^{n+1}$ with~(\ref{eq:proposal}).
\item If the following energy bound does not hold
  \begin{equation}
    \label{eq:energy-bound}
    \min\left(\varepsilon_i^n,\varepsilon_j^n\right) > \frac{m}8\left([\vect{v}_{ij}^{n+1}]^2-[\vect{v}_{ij}^n]^2\right),
  \end{equation}
  the move is rejected: $(\vect{p}_i^{n+1},\vect{p}_j^{n+1},\varepsilon_i^{n+1},\varepsilon_j^{n+1})=(\vect{p}_i^n,\vect{p}_j^n,\varepsilon_i^n,\varepsilon_j^n)$.\\
  If the bound is satisfied, the algorithm continues.
\item Compute the acceptance ratio with~\eqref{eq:metropolis-ratio}.
\item Draw $U_{ij}^n \sim \mathcal{U}[0,1]$ and compare it with $A_{\Delta t}(\vect{v_{ij}^n},\vect{v}_{ij}^{n+1})$. If $U_{ij}^n > A_{\Delta t}(\vect{v_{ij}^n},\vect{v}_{ij}^{n+1})$, the move is rejected and $(\vect{p}_i^{n+1},\vect{p}_j^{n+1},\varepsilon_i^{n+1},\varepsilon_j^{n+1})=(\vect{p}_i^n,\vect{p}_j^n,\varepsilon_i^n,\varepsilon_j^n)$.\\
  Otherwise it is accepted and the momenta and internal energies are updated with~\eqref{eq:proposal-update}, along with the internal temperatures $T_i$, $T_j$ and heat capacities $C_i$, $C_j$. 
\end{enumerate}

\subsubsection{Approximate Metropolized scheme}
\label{sec:approx-met}

Since the computation of the Metropolis ratio may be cumbersome in practical simulations, we propose a simplified and approximate scheme where we only reject moves that cause internal energies to become negative.
It avoids the need to actually compute the Metropolis acceptance ratio.

As for the complete Metropolized scheme, we use the expression~(\ref{eq:proposal}) as the proposed evolution for the relative velocities.
We then check whether the updated internal energies remain positive and reject the moves that do not satisfy this property.
The current configuration at time is then used as the new configuration and counted as usual in the averages.
Otherwise the move is accepted and the velocities and internal energies, along with the internal temperatures and heat capacities, are updated accordingly.
When no stability issues, \emph{i.e.} negative internal energies, appear, the Approximate Metropolis Scheme (AMS) is equivalent to SSA.

\section{Numerical results}
\label{sec:results}

\begin{figure}[!ht]
  \centering
  \includegraphics{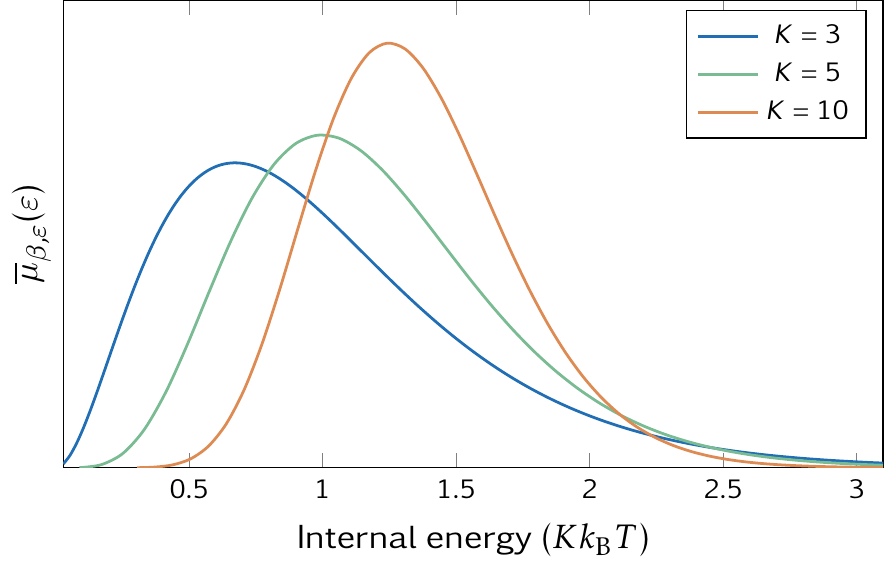}
  \caption{Distribution of internal energies (in units of $Kk_{\rm B}T$) with the ideal gas equation of state.}
  \label{fig:distribution-k}
\end{figure}

In the following, we test the accuracy of our schemes for the ideal gas equation of state given by
\begin{equation}
  \label{eq:pg-eos}
  \mathcal{S}_{\rm ideal}(\varepsilon,\rho) = \frac32(K-1)k_{\rm B}\ln(\varepsilon) - \frac12(K-1)\ln(\rho).
\end{equation}
The interest of this model is that the marginal distribution for the internal energies $\varepsilon_i$ has an analytic expression:
\begin{equation}
  \label{eq:dist-analytic}
\overline{\mu}_{\beta,\varepsilon}(\rd \varepsilon) = \frac{ \beta^{ \frac{C_K}{k_{\rm B}}} }{ \Gamma\left(\frac{C_K}{k_{\rm B}}\right) }\varepsilon^{\frac{C_K}{k_{\rm B}}-1}\exp\left(-\beta\varepsilon\right)\,\rd\varepsilon,
\end{equation}
where $C_K = \frac32(K-1)k_{\rm B}$ is the heat capacity in the equation of state~\eqref{eq:pg-eos} and $\Gamma$ is the Gamma function.
This distribution is plotted in Figure~\ref{fig:distribution-k} for various particle sizes.
As the size $K$ decreases, very small internal energies become more likely and stability issues arise.

Our simulations have been carried out on a 3-dimensional system of 1000 particles initialized on a simple cubic lattice with an initial temperature $T = 1000$~K.
The internal energies are chosen so that $T_i(\varepsilon_i,\rho_i(\vect{q})) = T$ with the density $\rho_i(\vect{q})$ evaluated from the initial distribution of the positions; while the velocities are distributed along the Boltzmann distribution.
We let the system equilibrate during a time $\tau_{\rm therm} = 50$ to obtain an equilibrated initial configuration.
The shear viscosity is set to $\eta = \num{2e-3}$~Pa.s and we neglect the bulk viscosity $\zeta$.

\subsection{Integrating the fluctuation/dissipation dynamics}
\label{sec:results-fd}

We first investigate the properties of the integration schemes for the fluctuation/dissipation part only and do not couple SSA and the Metropolized schemes with Velocity Verlet.
While SSA is quite stable for large particles ($K>10$) for which time steps as large as $\Delta t = 5$ can be used with no occurrence of a negative internal energy during a simulation time $\tau_{\rm sim}$, stability issues arise for smaller particles.
At $K=5$, we need a time step $\Delta t < 0.025$ to avoid the appearance of negative internal energies with SSA. As a comparison, the stability limit for the Verlet scheme at $K=5$ is $\Delta t = 0.8$.
As the particle size decreases further, it becomes impossible to run simulations and no admissible time step has been found for $K=2$.
With the rejection of moves provoking negative energies, the Metropolized schemes are stable at any time step for every particle sizes.

\begin{figure}[!ht]
  \centering
  \subfigure[Distributions at $\Delta t =1$.]{\includegraphics[width=.45\columnwidth]{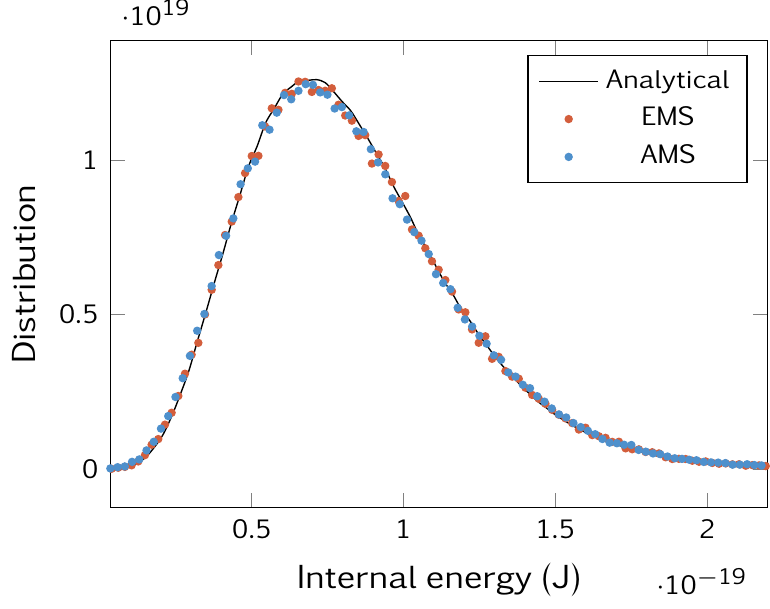}\label{fig:dist-dt-1}}
  \subfigure[Bias with respect to the time step]{\includegraphics[width=.45\columnwidth]{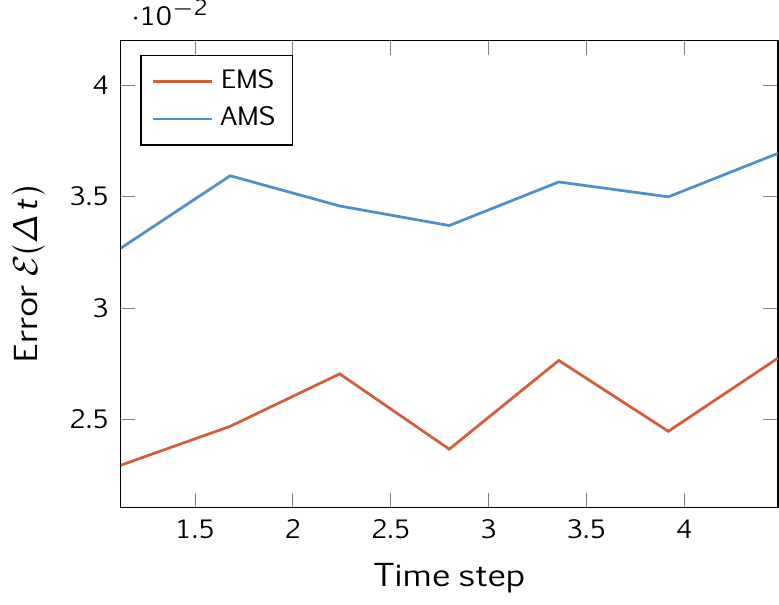}\label{fig:dist-bias}}
  \caption{Comparisons of the internal energy distributions with the ideal gas equation of state for the exact (EMS) and approximate (AMS) Metropolized algorithms: for $\Delta t =1$ in~\protect\subref{fig:dist-dt-1} and with the error~\eqref{eq:error-dist} with respect to the time step in~\protect\subref{fig:dist-bias}}
  \label{fig:distribution-dt}
\end{figure}

When they are not coupled to Velocity Verlet, the SSA and Metropolized schemes preserve exactly the energy by construction.
We can however compare the bias in the distributions of internal energies for the different schemes.
Figure~\ref{fig:distribution-dt} shows the distributions of internal energy for the exact and approximate Metropolized scheme using the ideal gas equation of state~(\ref{eq:pg-eos}) with $K=5$ obtained with a simulation time $\tau_{\rm sim}=20000$, compared with the analytic distribution~\eqref{eq:dist-analytic}.
In practice, the distributions $\nu_{\Delta t}$ obtained from the numerical simulations are approximated using histograms computed on $50$ configurations extracted at regular time intervals from the simulations.
This ensures a constant number of sampling points for all time steps.
The histograms consist in $N_{\rm bins} = 150$ bins uniformly distributed between $\varepsilon_{\rm min} = 0.1k_{\rm B}T$ and $\varepsilon_{\rm max} = 20k_{\rm B}T$.
A more quantitative measurement of the bias is to evaluate the quadratic error with respect to the theoretical distribution
\begin{equation}
  \label{eq:error-dist}
  \mathcal{E}(\Delta t) = \sqrt{\frac{\int_0^{+\infty} \left[\nu_{\Delta t}(\varepsilon)-\overline{\mu}_{\beta,\varepsilon}(\varepsilon)\right]^2~\rd \varepsilon}{\int_0^{+\infty}\overline{\mu}_{\beta,\varepsilon}(\varepsilon)^2\,\rd \varepsilon}}.
\end{equation}
Due to the Metropolis procedure, the only source of error for EMS is of statistical nature.
This is not guaranteed for AMS but no systematic bias is apparent up to $\Delta t = 5$.
The agreement with the theoretical distribution is a bit deteriorated when the full acceptance ratio is not computed and AMS displays a $30\%$ larger error compared to the exact Metropolization.

\begin{figure}
  \centering
  \subfigure[EMS (Total)]{\includegraphics[width=.3\columnwidth]{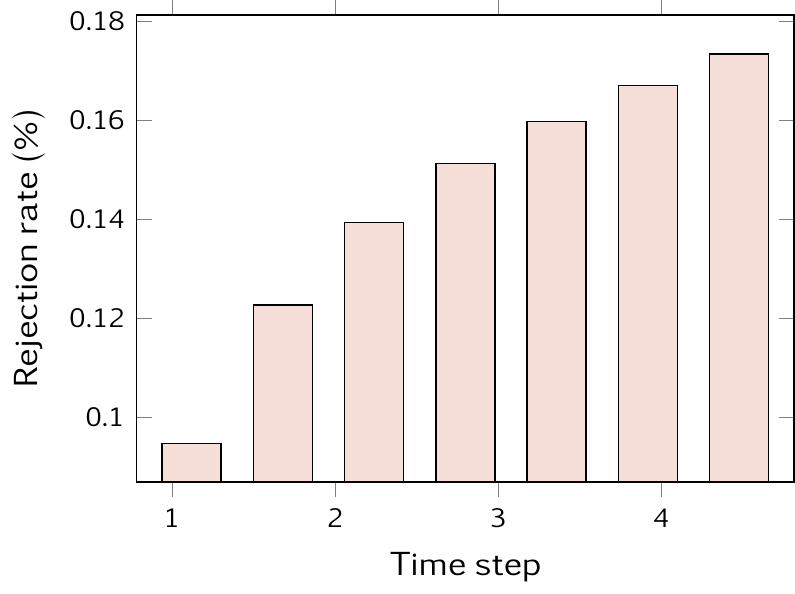}\label{fig:rejection-met}}
  \subfigure[EMS (Negative energies)]{\includegraphics[width=.3\columnwidth]{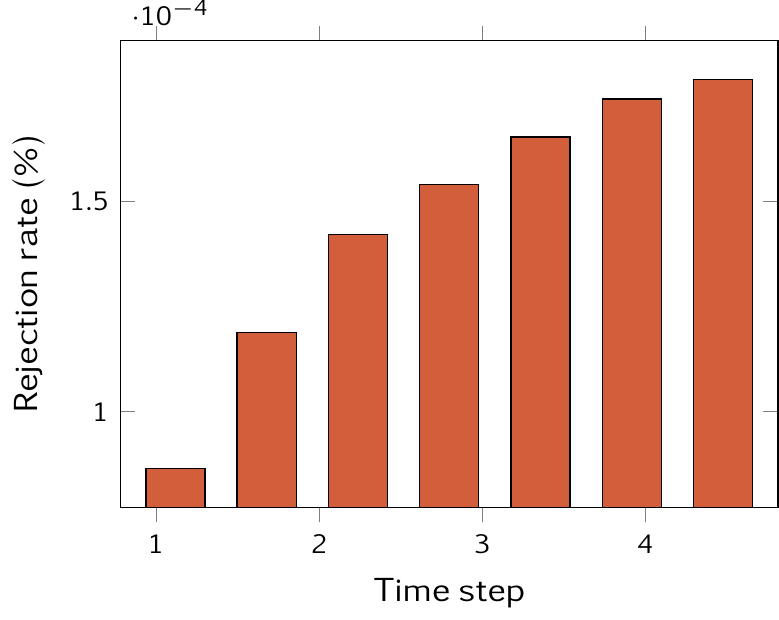}\label{fig:rejection-met-neg}}
  \subfigure[AMS]{\includegraphics[width=.3\columnwidth]{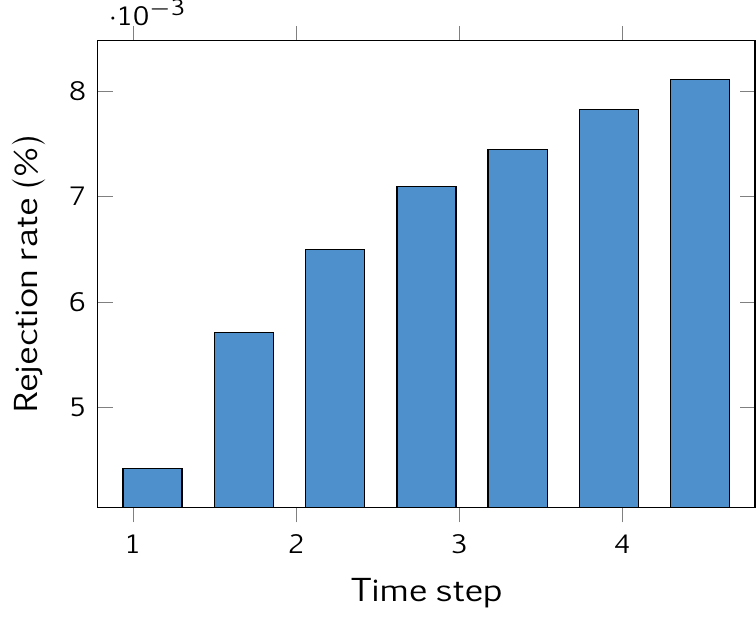}\label{fig:rejection-am}}
  \caption{Rejection rates for~\protect\subref{fig:rejection-met} the exact Metropolization and~\protect\subref{fig:rejection-am} the approximate Metropolization. The rejection rate due to negative internal energies in the exact Metropolis scheme is also displayed in~\protect\subref{fig:rejection-met-neg}.}
  \label{fig:rejection}
\end{figure}

We also observe the scaling of the rejection rate with the time step in Figure~\ref{fig:rejection}.
The Metropolized scheme displays rejection rates between $0.1\%$ and $0.2\%$ for time steps between $\Delta t = 1$ and $\Delta t = 5$.
For our system size ($1000$ particles), it means that in average several fluctuation/dissipation interactions are rejected each time step.
Most of these rejections are due to the Metropolis ratio and not to the appearance of negative internal energies which accounts for approximately one rejection every thousand.
When we only reject forbidden moves that would cause negative energies, the rejection rate of this approximate Metropolization is between $0.005\%$ and $0.01\%$, which is about $20$ times smaller than the overall rejection rate of the exact Metropolization.
This is however two orders of magnitude larger than the occurrence of negative internal energies with the exact Metropolis scheme.
By rejecting authorised but unlikely moves (leading to small energies for instance), EMS is less prompt to the apprition of negative internal energies.
A linear fit in log scale shows that the total rejection rate for both the exact and approximate Metropolization scales as $\Delta t^{0.42}$.
The rejection of negative energies with the exact Metropolis schemes roughly follows the same scaling with $\Delta t^{0.5}$.

\subsection{Integrating the full SDPD}
\label{sec:results-full}

We now turn to the numerical integration of the full dynamics and study the behavior of the full schemes coupling the Velocity Verlet scheme for the integration of the conservative part of the dynamics and either SSA or its Metropolized versions for the fluctuation/dissipation dynamics.

\begin{figure}[!ht]
  \centering
  \includegraphics{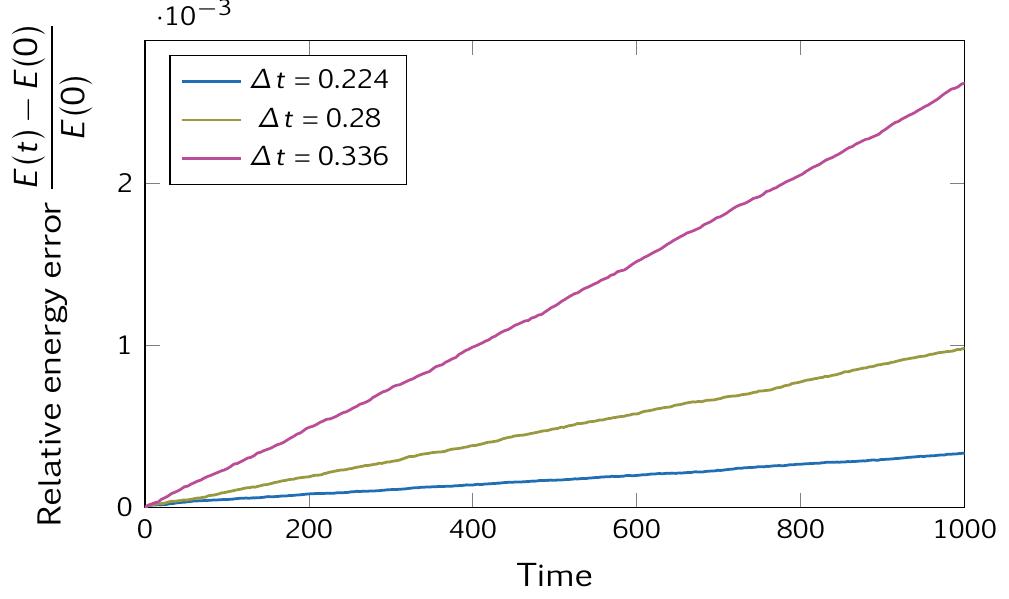}
  \caption{Time evolution of the total energy with the Metropolized algorithm for different time steps. A linear drift in the energy is observed as is usual in such methods.}
  \label{fig:energy-time-met}
\end{figure}

\begin{figure}[!ht]
  \centering
  \includegraphics{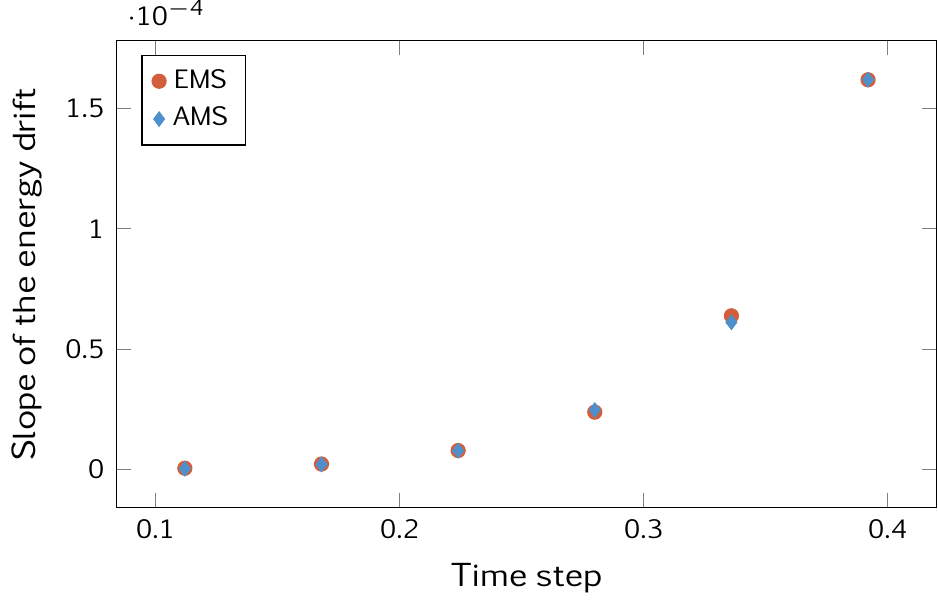}
  \caption{Slope of the energy drift with respect to the time step for the two Metropolized schemes (EMS and AMS)}
  \label{fig:energy-drift}
\end{figure}

\begin{figure}[!ht]
  \centering
  \includegraphics{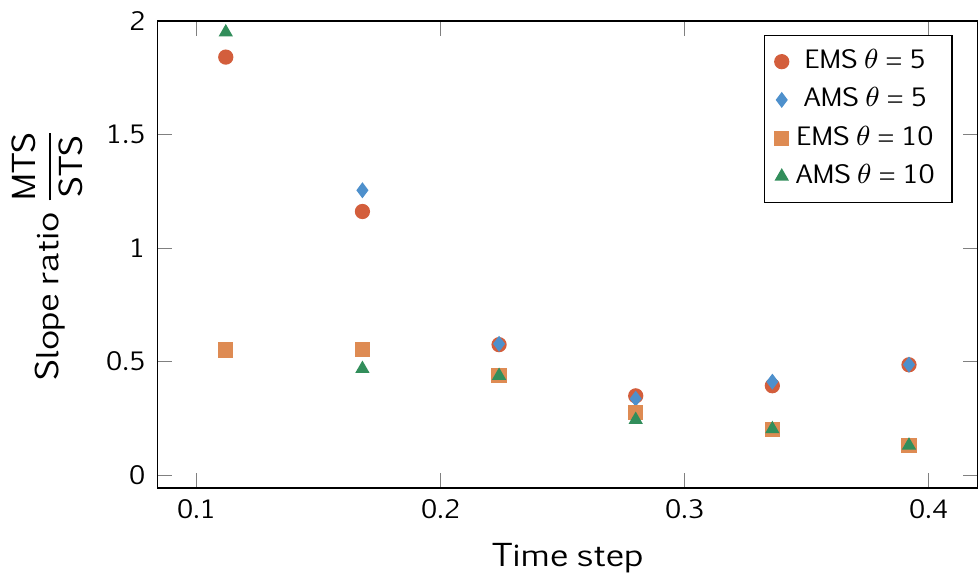}
  \caption{Ratio between the slope of the energy drift for the multiple time steps Metropolized schemes (EMS and AMS) and their STS versions (with $\Delta t = \Delta t_{\rm VV}$). }
  \label{fig:ratio-mts}
\end{figure}

To evaluate the energy conservation and the scheme stability, we run $n_{\rm sim}=10$ independent simulation with $K=5$ during a time $\tau_{\rm sim}=1000$ for each time step $\Delta t$ and average the results.
The schemes obtained by superimposing~\eqref{eq:sdpd-verlet} and either SSA or the Metropolized scheme lead to linear energy drifts which have already been observed in DPDE~\cite{lisal_2011,homman_2016} and in SDPD~\cite{faure_2016}.
This is illustrated in Figure~\ref{fig:energy-time-met} where the total energy with respect to time is displayed for different time steps in the case of the Metropolized scheme.
We characterize the energy drift by fitting the time evolution of the energy on a linear function and plot the resulting slope  in Figure~\ref{fig:energy-drift} for SSA, Metropolis and its approximate version.
We observe a similar energy drift for all methods.
As in the integration of the fluctuation/dissipation only, SSA is limited to small time steps to prevent stability issues to arise while the Metropolized schemes greatly increases the admissible time steps.
However, the time steps for which the dynamics is stable are much smaller with the full dynamics than those reported in Section~\ref{sec:results-fd}.
Although the conservative interactions are bounded in SDPD unlike the DPDE simulations in~\cite{stoltz_2017}, they still induce a stringent stability limit on the time steps.
This observation leads us to consider multiple time step implementations (MTS) where the fluctuation/dissipation is integrated with a larger time step.
We introduce the time steps $\Delta t_{\rm VV}$ used to integrate the conservative part with the Velocity-Verlet scheme and $\Delta t_{\rm FD} = \theta\Delta t_{\rm VV}$ used for the discretization of the fluctuation/dissipation with a Metropolized scheme (EMS or AMS).
We test this approach with $\theta = 5$ and $\theta = 10$.
The algorithm then reads:
\begin{enumerate}
\item $\theta$ consecutive steps of Velocity Verlet with $\Delta t = \Delta t_{\rm VV}$.
\item One step of EMS or AMS with $\Delta t = \theta\Delta t_{\rm VV}$.
\end{enumerate}
We plot in Figure~\ref{fig:ratio-mts} the slope of the energy drift compared to their single time step (STS) version (with a time step $\Delta t = \Delta t_{\rm VV}$).
For both the exact and the approximate Metropolization, the energy drift rate is smaller for the multiple time step approach when the time step is large enough.
Moreover, the reduction of the energy drift is enhanced for larger $\theta$ with a division by $6$ of the rate for $\theta = 10$ and $\Delta t = 0.392$ when it is only halved for $\theta = 5$.

\begin{table}[!ht]
  \centering
  \begin{tabular}{cc}
    \toprule
    Scheme & Time ($\mu$s/iter/part)\\
    \hline
    Verlet & 37.66\\
    SSA & 49.48\\
    EMS & 63.39 \\
    AMS & 51.07 \\
    EMS with MTS & 43.57\\
    AMS with MTS & 40.92\\
    \bottomrule
  \end{tabular}
  \caption{Comparison of the computation time per iteration per particle for the Shardlow-like algorithm and the Metropolis schemes. For the multiple time step implementations, the time steps ratio is set to $\theta = 5$. The time for Velocity Verlet only is given as a reference.}
  \label{tab:schemes-time}
\end{table}
We measure the time per iteration and per particle for the different Metropolis schemes with $\Delta t=\num{2.24e-2}$ or $\Delta t_{\rm VV}=\num{2.24e-2}$ and gather them in Table~\ref{tab:schemes-time}.
For the multiple time step algorithms, the number of iteration is the number of Verlet steps (which is thus the same than in the STS case).
The integration of the fluctuation/dissipation dynamics with SSA represents a quarter of the total computational time.
The integration of the fluctuation/dissipation part with the exact Metropolization is about twice as long since we need to compute the reverse move and estimate the Metropolis ratio.
This results in an overall increase by $30\%$ of the total simulation time.
However, much larger time steps can be chosen with EMS while SSA suffers from stringent stability limitations.
There is almost no overhead when resorting to the approximate Metropolization which also greatly improves the stability and is as good as EMS in terms of energy conservation.
With the multiple time step strategy, the time needed for the fluctuation/dissipation is greatly reduced, as expected, by a factor $\theta$.

\subsection{Simulation of nonequilibrium systems}
\label{sec:results-noneq}

While all the previous simulations were carried in an equilibrium situation, the Metropolization procedure we propose are also suited for nonequilibrium settings.
In particular, SDPD has been applied to model shock waves~\cite{faure_2016} and reactive waves~\cite{faure_2017c}.
Stability is a crucial issue for these phenomena since they involve dramatic changes in the thermodynamic states of the material.
We thus illustrate the enhanced stability of the Metropolized schemes in non equilibrium situations by simulating a shock wave.
The system is initialized as previously mentioned but with $N=23400$ particles organized on $10 \times 10 \times 234$ lattice with periodic boundary conditions in the $x$- and $y$-directions.
In the $z$-direction, two walls are located at each end of the system and formed of ``virtual'' SDPD particles as described in~\cite{bian_2012,faure_2016}.
These virtual particles interact with the real SDPD particles through the conservative forces~(\ref{eq:cons-forces}) and a repulsive Lennard-Jones potential that ensures the impermeability of the walls.
After the system equilibration during $\tau_{\rm therm} = 50$, the lower wall is given a constant velocity $v_{\rm P} = 1661$~m.s$^{-1}$ in the $z$-direction.

\begin{figure}[!ht]
  \centering
  \includegraphics{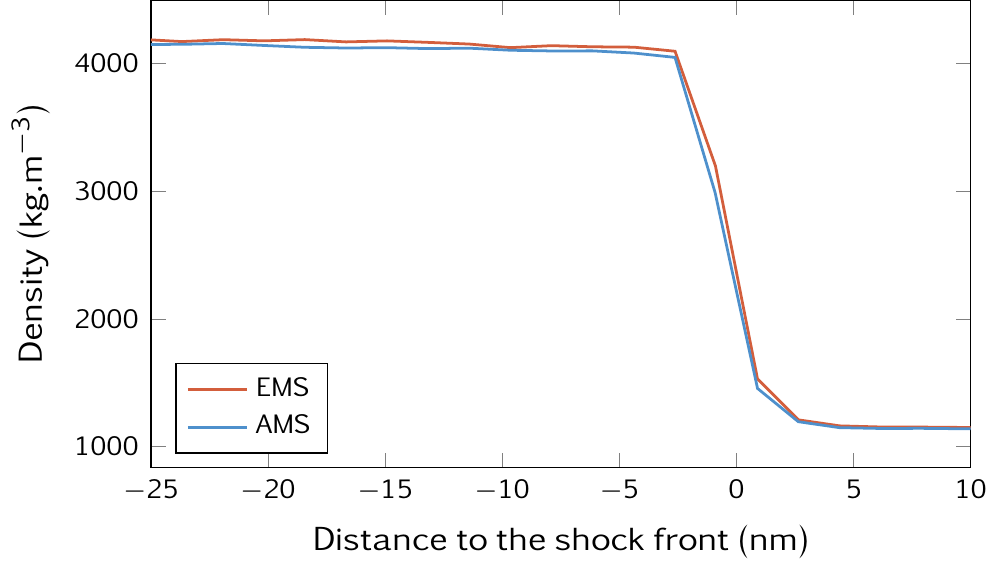}
  \caption{Density profile in the reference frame of the shock front for the exact and approximate Metropolization.}
  \label{fig:shock-profiles}
\end{figure}

To obtain the profiles of physical properties at a given time, the simulation box is divided into $n_{\rm sl}=100$ slices regularly distributed along the $z$-axis in which the physical properties are averaged.
Since a shock wave is a stationary process in the reference frame of the shock front, we can average profiles over time after shifting the position of the shock front to $z=0$.
We plot in Figure~\ref{fig:shock-profiles} the density profile for the Metropolized schemes (EMS and AMS) with $\Delta t = 0.045$.
This choice is governed by the piston velocity $v_{\rm P}$ and ensures than the piston does not move by more than $20\%$ of the characteristic distance between two particles.
This avoids instabilities in the conservative part of the dynamics.
The Rankine-Hugoniot relations makes use of the conservation of mass, momentum and energy to predict the thermodynamic properties in the shocked state from the initial thermodynamic conditions and the velocity of the particles in the shocked region.
The properties estimated from the SDPD simulations, namely the velocity of the shock wave $v_S$, the density $\rho_s$, pressure $P_s$ and temperature $T_{S}$ in the shocked state, all agree very well with the theoretical predictions as can be seen in Table~\ref{tab:shock-properties}.
\begin{table}[!ht]
  \centering
  \begin{tabular}{ccccc}
    \toprule
    Scheme & $v_S$ (km.s$^{-1}$)& $\rho_S$ (kg.m$^{-3}$)& $P_S$ (GPa)& $T_S$ (K)\\
    \hline
    EMS & 2254 & 4173 & 4.50 & 7816\\
    AMS & 2268 & 4151 & 4.49 & 7836\\
    RH & 2314 & 4075 & 4.58 & 8244\\
    \bottomrule
  \end{tabular}
  \caption{Average observables in the shocked state: SDPD compared to the Rankine-Hugoniot (RH) predictions.}
  \label{tab:shock-properties}
\end{table}
Let us point out that this simulation would not have been possible with SSA since negative internal energies appear very early in the simulation even for time steps as small as $\Delta t = 10^{-4}$.
The Metropolization procedure can thus be particularly useful in nonequilibrium simulations where stability issues are aggravated.

\section{Conclusion}
\label{sec:conclusion}

In this work, we have introduced a Metropolis procedure for the integration of the fluctuation/dissipation part in SDPD.
This adaptation of the Metropolized schemes for DPDE has led to a significant increase of the stability of the dynamics for small particle sizes.
This allows us to carry simulations that traditional schemes could not achieve due to the very stringent limitation on the time step they must respect.
It appears that an approximate version of the Metropolis step where only the negative internal energies are rejected is actually enough to ensure the stability of the algorithm and does not display a larger bias or energy drift than its exact version.

With the addition of the Metropolis step, the integration of the fluctuation/dissipation is stable for very large time steps and the limit on the admissible time steps emerges from the conservative part. A multiple time step approach has been tested where a smaller time step was used for Velocity Verlet and a larger one for the Metropolized SSA scheme.
This resulted in a similar energy drift at a reduced computational cost.

The relevance of the Metropolization in nonequilibrium situations has been illustrated by the simulation of a shock wave.
While traditional schemes such as SSA fail to perform the simulation for small particles, due to the aggravated stability issues, both the exact and approximated Metropolized schemes have allowed us to recover the correct physical properties of the shock wave.

Our Metropolis schemes still suffer from the difficult parallelization of SSA on which they are based.
It would be most beneficial to adapt a stabilization procedure based on rejecting moves leading to negative energies to parallel schemes~\cite{larentzos_2014,homman_2016} in order to deal with larger systems.

\section*{Acknowledgments}
We thank J.-B. Maillet for helpful discussions.
The work of G.S. was funded by the Agence Nationale de la Recherche, under grant ANR-14-CE23-0012 (COSMOS) and by the European Research Council under the European Union's Seventh Framework Programme (FP/2007-2013) / ERC Grant Agreement number 614492.

\end{document}